Running head: AESTHETICS AND ASTRONOMY

Aesthetics and Astronomy: Studying the public's perception and understanding

of imagery from space

Lisa F. Smith and Jeffrey K. Smith

University of Otago

Kimberly K. Arcand, Randall K. Smith, and Jay Bookbinder

Harvard-Smithsonian Center for Astrophysics

Kelly Keach

University of Otago

Abstract

This study examined the scientific understanding of astronomical pictures by experts and non-experts. It explored how both groups perceive astronomical images, and their understanding of both what they are seeing and the science that underlies the images. Data comprised $n = 8866$ responses from a survey linked to the NASA Astronomical Picture of the Day web site and 4 focus groups held at the Harvard-Smithsonian Center for Astrophysics. Results indicated that variations in presentation of color, explanation, and scale affect comprehension of astronomical imagery. With those who are more expert, shorter, more technical explanations are effective; however, scales are still necessary for complete comprehension. Experts tend to look at the images from a more scientific, data-oriented perspective while non-experts are more likely to focus, especially initially, on the aesthetic or emotional values of the images. Results provide suggestions for educational outreach to the public.

Aesthetics and Astronomy: Studying the public's perception and understanding

of non-traditional imagery from space

Some 400 years after Galileo, modern telescopes have enabled humanity to "see" what the natural eye cannot. Astronomical images today contain information about incredibly large objects located across vast distances and reveal information found in "invisible" radiation ranging from radio waves to X-rays. The current generation of telescopes has created an explosion of images available for the public to explore. This has, importantly, coincided with the maturation of the Internet. Every major telescope has a web site, often with an extensive gallery of images. New and free downloadable tools exist for members of the public to explore astronomical data and even create their own images.

In short, a new era of an accessible universe has been entered, in which the public can participate and explore like never before. But there is a severe lack of scholarly and robust studies to probe how people – especially non-experts – perceive these images and the information they attempt to convey. Most astronomical images for the public have been processed (e.g., color choices, artifact removal, smoothing, cropping/field-of-view shown) to strike a balance between the science being highlighted and the aesthetics designed to engage the public. However, the extent to which these choices affect perception and comprehension is, at best, poorly understood. The goal of the studies presented here was to begin a program of research to better understand how people perceive astronomical images, and how such images, and the explanatory material that accompanies them, can best be presented to the public in terms of understanding, appreciation, and enjoyment of the images and the science that underlies them.

The Perception of Images

Because the study of the perception of astronomical images is in its infancy, it is not possible to draw directly on prior research. As a starting point, therefore, we look to the scholarly literature in three areas that we believe can inform this research: museum behavior and learning, aesthetics, and expert/novice differences. Although astronomical images are primarily viewed on the Internet or as part of classroom instruction, as opposed to in museums, the research on museums provides us with a source of information on how people view similar objects such as works of art, science displays, and historical and archaeological artifacts. With regard to museum behavior and learning, one of the fundamental issues is that museum visitation is voluntary in nature (Falk, 2005). Except for school groups, the individuals in museums are there because they want to be. And they attend to the displays and exhibitions that interest them, staying as long as that interest is maintained (Rounds, 2004). In a museum setting, the desire to educate needs to be mediated with the inherent interest levels and background of the visitors (Smith, 2008; Smith & Smith, 2006). In a similar fashion, it is important to understand the voluntary, self-directed viewing of astronomical images on the Internet (and, when applicable, astronomical images in museum settings).

The voluntary nature of participation occurs not only at the level of the museum as a whole or of an exhibition, but also at the level of the individual object or image. Smith and Smith (2001) found that, on average, visitors to The Metropolitan Museum of Art spent 27 seconds in front of masterpiece paintings (e.g., Velasquez' "Juan de Pareja;" Rembrandt's "Aristotle Contemplating the Bust of Homer"). Participants in their *in situ* observational study never spent more than 4 minutes looking at a single work of art. Although this research was conducted in a museum, the findings speak to the issue at looking at images online as well, which is an important component of looking at astronomical images. For example, NASA's Astronomy Picture of the Day web site alone receives up to a million page views a day (J. Bonnell, co-originator and editor of the APOD website, 2009, personal communication). Visitors to this site, however, are typically at home, school, or work when visiting. They are not in a museum, but may be simply taking a break from their daily routines. Their visits are typically brief. Thus, our first issue, both with regard to visits to an exhibition or a museum, and to looking at individual images (or small sets of them) online, has to do with the "holding power" of the works under consideration. If people do not spend time on the works, it is difficult for there to be a serious level of appreciation or learning. Thus, we are interested in how different aspects of presentation of an image influences how long people look at those images.

Of course, what people get out of viewing an image is dependent not only on how long they look at the image, but what they do while looking at the image, and what kind of information, if any, accompanies the image (Russell, 2003). The second issue under consideration in this research concerns the effect of providing information about images to viewers, and how best to present that information. There are a number of studies that have shown that what people do in museums affects their understanding and appreciation of the works. People talk with others (Ballatyne & Packer, 2005; Smith & Smith, 2003), read wall texts and object labels (Smith & Carr, 2001), take audio tours (Tinio, 2005), and return to look at objects again (Smith & Smith, 2003). Related to considerations of visitor behavior in museums is the question of what kind of information is presented to accompany the object. This has been a subject of study for a number of years. Cupchik, Shereck, and Spiegel (1994) found that information that helped people interpret artwork enhanced their sense of understanding and meaning in those artworks. Russell and Milne (1997) reported a similar finding, but Millis (2001) cautioned that a balance needs to be struck between providing adequate information and inundating the viewer. Two of the authors of this study, over a career of conducting research on people in museums, have found that people want both more and less information at the same time (J. Smith, 2009). That is, museum visitors have reported that they find long labels accompanying works of art off-putting, while simultaneously saying that they want more information about those same works. This phenomenon is quite strong, and provides a dilemma for those who wish to communicate to the public about images. Pekarik (2004) encapsulated this issue nicely in a piece called, "To explain or not to explain."

The approach to conducting the research presented here draws substantially from work in the psychology of aesthetics. Aesthetics is a field that is perhaps as old as philosophy itself, and the psychology of aesthetics is certainly as old as psychology, with the fathers of modern psychology, Fechner (1871) and Wundt (1874), being very much interested in what we find beautiful. Arnheim (1954) and Berlyne (1974) pioneered a revival of interest in the psychology of aesthetics that carries on to today. Recent work in the psychology of aesthetics has focused on the reactions of individuals to works of art (Locher, Smith, & Smith, 2001), and the development of models for how individuals understand works of art (Leder, Belke, Oeberst, & Augustin, 2004). There is a focus today on more cognitive approaches to understanding how individuals visually process art. These models incorporate the background information and affect that the individual brings to the work, the nature of the work itself, and the explanatory information that accompanies the work. The work is viewed as a kind of problem, or challenge, that the viewer has to solve. Perhaps more precisely, the viewer uses the work as a basis for posing a personal challenge to him or herself. What is this work about?  Do I like it or not?  What is it telling me?  The level of success of the encounter with the work is also personal, and influences the cognitive and affective outcomes of the encounter (Arcand et al., 2009: Leder, et al. 2004; L. Smith, 2009; Smith, Smith, Arcand, Smith, & Holterman ten Hove, 2009). This approach to viewing images may provide a useful framework for helping to understand how individuals view astronomical images as well. The astronomical image contains both aesthetic and scientific content; our reactions to astronomical images typically encompass the affective as well as the cognitive. Therefore, research on the psychology of aesthetics informs both the designs and the approaches to measurement used in this study.

The third area of research that we draw upon has to do with expert/novice differences. Research in expert/novice differences began in the early 1980s (Chi, Feltovich, & Glaser, 1981) and has been extended to a wider variety of domains including differences in chess players, engineers, schoolteachers, and people looking at art (Cook, 2006; Nodine, Locher, & Krepinski, 1993; Silvia, 2006; Simon, 1990). One of the more robust findings of this literature is that experts have a store of knowledge that they can call upon without using extensive cognitive resources. Thus, an expert chess player can immediately see that a board in play represents certain approaches and defenses and knows what options are available and most likely to lead to successful play. In contrast, a novice player has to plod through numerous possibilities and look carefully at the placement of all pieces on the board, making the generation of good alternatives extremely difficult. Similar findings have been observed repeatedly throughout this literature: experts, with a vast amount of experience to draw upon, see situations and images in fundamentally different ways than novices (e.g., Chi, Feltovich, & Glaser, 1981; Chi, Glaser, & Rees, 1982;Cook, 2006; diSessa, 2004; Larkin, McDermott, Simon,& Simon, 1980; Patrick, Carter, & Wiebe, 2005; Simon, 1990). The question arises as to whether such expert/novice differences are found in the perception of astronomical images.

The Present Study

To address these and related questions, in 2008 The Aesthetics and Astronomy Group

(A&A) was formed. A&A is a collaboration of astrophysicists, astronomy image development professionals, educators, and specialists in the aesthetic and cognitive perception of images. A&A was formed specifically to examine how astronomical images are perceived and how they communicate science to the public.

We present here the results of two studies looking at these issues. The first study was a randomized experimental design using images on computer screens. The intent was to study the public's perception of multi-wavelength astronomical imagery and the effects of the scientific and artistic choices in processing astronomical data and compare their perceptions to those of professional astrophysicists. The second study was a qualitative study based on a series of focus groups, conducted at the Harvard-Smithsonian Center for Astrophysics in Cambridge, Massachusetts.

The research questions for this study were:

1. How much do variations in presentation of color, explanatory text, and illustrative scales affect comprehension of, aesthetic attractiveness, and time spent looking at deep space imagery?

2. How do novices differ from experts in terms of how they look at astronomical images?

Method

The first study was based on a large sample of individuals who volunteered to participate in our research via an invitation placed on several popular astronomy web sites. This study involved responding to images in an online survey and being assigned randomly to various experimental conditions concerning text, color, and the use of illustrative scales. We refer to this study as the web-based survey/experiment. The second study was a series of four focus groups held at the Harvard-Smithsonian Center for Astrophysics. We refer to this as the focus group study. The combination of the two approaches allowed us to obtain experimental rigor in looking at how individuals responded to various approaches to presenting images, scales for interpretation of the images, and text to enhance understanding of the images. The focus group study let us explore in depth just how people go about processing the images that they see, and to look carefully at expert/novice differences in this processing. We present the methods for the web-based experiment first, followed by the methods for the focus group.

*Study 1:  The Web-based Experiment*

In the web-based survey/experiment there were $n = 8866$ useable protocols from a web-based survey (described below). Respondents represented 103 countries, with the majority from the USA ($n = 5562$), the UK ($n = 555$), and Canada ($n = 541$). Males ($n = 6966$) responded more frequently than females ($n = 1900$) by over 3:1. The ages of the participants were evenly distributed from 15-64 years; $n = 874$ respondents reported their age as over 65. Participants rated their level of knowledge about astronomy on a 1 (complete novice) to 10 (expert) scale. The ratings were $n = 468$ (1), $n = 908$ (2), $n = 1519$ (3), $n = 1253$ (4), $n = 1246$ (5), $n = 1159$ (6), $n = 1175$ (7), $n = 648$ (8), $n = 260$ (9), and $n = 218$(10), with $n = 12$ not reported.

*Materials*

A&A members created a web-based survey that asked respondents to evaluate astronomical images. The survey contained a section of demographic items followed by a set of items using astronomical images, described in the Procedure and Results sections.

*Procedure*

For one week in Autumn 2008, an invitation to participate in the web-based survey/experiment was advertised on well-trafficked, public-friendly web sites such as NASA's Astronomy Picture of the Day (APOD, http://apod.nasa.gov/) and the Chandra X-ray

Observatory web site (http://chandra.harvard.edu). These web sites featured a box with an explanation of this project and an invitation to participate in an online survey. Clicking on the box led to the first page of the web-based survey/experiment. The URL was also circulated to schools, universities, and other groups and individuals, with encouragement to forward the information to any who might be interested.

The web-based survey/experiment began with an introductory page describing the study and containing an informed consent protocol. A control was included to prevent multiple responses from the same computer address. Following the introductory page, participants were given a short list of demographic items: age, gender, highest level of education, self-rating of expertise in astronomy on a 1 (complete novice) to 10 (expert) scale, residence, and familiarity with the APOD web site (which was expected to, and did, generate the largest number of responses).

Each participant then accessed the first of a series of items, each depicting an astronomical image. The images were presented with text or no text, variations in color, background imagery or no background imagery, and/or the presence or absence of scales. Participants were re-randomized for assignment to conditions for each item in the survey. After viewing an image with or without accompanying text, participants responded to various questions. The deep space images were taken from the Chandra, Hubble, Spitzer, GALEX, Very Large Array (VLA), and Hinode satellites and telescopes, among others, and included G292.0+1.8, M33 X-7, Whirlpool Galaxy (M51), and NGC 4696, and the Sun's poles. Latencies were collected to see how the experimental conditions influenced viewing times.

*Analysis*

The web-based survey/experiment data involved random assignment to various conditions, conducted anew for each of the items in the survey. Thus, this was in essence a series of experiments conducted on the same sample. Given the large sample size, we chose an alpha level of .01 for statistical significance, and used a Bonferroni adjustment to account for multiple statistical tests being made. Analysis of variance and *chi* square techniques were used to analyze the data. We occasionally used analysis of variance with dichotomous outcome variables, as the very large sample size provided for sufficient variation and allowed for analyses to be comparable from one measure to another.

One of the variables examined was latency, or time spent looking at images. In examining the distributions of latencies, we noted that some individuals spent an unrealistic amount of time on some objects (up to several hours in some cases). It occurred to us that the data collection procedures included individuals who did not complete a task, but perhaps left their computers and did not return, or were interrupted during their participation in the study. After looking at the time distributions, we decided to truncate the most extreme 1% of the scores for all latency data, which effectively brought both tails of the distribution into a more normal spread.

Two types of effect sizes are reported in the analyses. When the difference between two means is discussed, Cohen's *d* was used as the estimate of the effect size. When more than two means are discussed, *eta* squared was used. No effect size was used when discussing the differences between simple percentages. All *post hoc* analyses were based on Sheffé procedures.

*Study 2:  The Focus Group Study*

In the focus group study, visitors were solicited through advertisements in local papers, online events calendars, and personal contacts. None of the focus group participants had participated in the web-based survey. There was a total of $n = 31$ volunteers for the focus groups: local high school students ($n = 8$), lay public ($n = 8$), local high school science teachers ($n = 6$), and astrophysicists and astronomers who work on image development ($n = 9$).

Semi-structured protocols were written to guide the discussion for the focus groups. The discussion questions centered on two images: Messier 101, and an image of the spicules on the pole of the Sun with superimposed scales intended to aid comprehension (see Appendix A). Participants also completed a paper-and-pencil activity with five images (see Appendix B). Packets containing a 2009 NASA calendar, DVD, bookmark, and additional educational and informational materials were assembled to thank the focus group participants.

*Procedure*

The focus group participants completed several activities in one 2-hour session per group. Using the two images, they responded to semi-structured items designed to explore how they viewed the image, their comprehension and misconceptions of the image and the underlying science associated with it, and their desires to learn more about the image and how it was made. The focus group of astrophysicists was also asked to discuss how they believe they communicate the science behind the images to non-experts.

*Analysis*

Focus group sessions were audio-recorded and transcripts were generated for each session. The transcripts were analyzed using NVivo 7.0 software for qualitative data analysis and a thorough review of the original transcripts by the researchers. We looked for themes within and across groups as well as areas of difference between experts and novices.

Results

The results of the web-based survey/experiment are presented first, followed by the results of the focus group study. The web-based survey/experiment addressed both research questions, concerning the influence of color, text and illustrative scales, and the differences between experts and novices. The focus group study primarily addressed expert/novice differences.

*The Web-based Survey/Experiment*

The results for the web-based survey/experiment are presented in order of the research questions. This approach involves a bit of repetition in the presentation of the analyses, but facilitates clarity in presenting the results. Furthermore, for the analysis of the first question, all of the participants were included in the analyses; for the expert/novice differences, we compared only those individuals indicating either very little knowledge of astronomy (self-rating of expertise = 1 or 2), or a great deal (self-rating of expertise = 9 or 10).

*The Influence of Color, Text, and Illustrative Scales*

The first research question concerned the influence of color, explanatory text, and illustrative scales on the comprehension, aesthetic attractiveness, and time spent looking at astronomical imagery.

*The Influence of Color Choice.* The first item on the survey that addressed this question is represented in Figure 1. This figure shows elliptical galaxy NGC 4696, a composite of X-ray, infrared and radio data. This item did not involve randomization. Rather, two versions of the same image were presented to all respondents, and they were asked to compare the images. The two images varied only in the choice of color palette; in one, regions of higher temperature were shown in red while in the other hotter elements were shown in blue. Respondents were asked which image they thought was more attractive, and which they thought was hotter in temperature. The sample as a whole had a slight preference for the "blue" image (53.3% to 46.7%), and they felt that the "red" image was hotter than the "blue" image (71.5% to 28.5%). The latter finding is particularly interesting as deep space images typically use blue to represent hotter (or more energetic) regions, and red to represent cooler (less energetic) areas. That being the case, the general public typically misperceives this message.

*The Influence of Text and Background Stars.* The second item involved a 2 X 2 experimental design, to examine how participants would react to an image with and without an explanatory text, and to the image being presented with and without a background of stars (see Figure 2). The image used for this experiment was the young Galactic supernova remnant, G292.0+1.8 in X-ray and optical light.

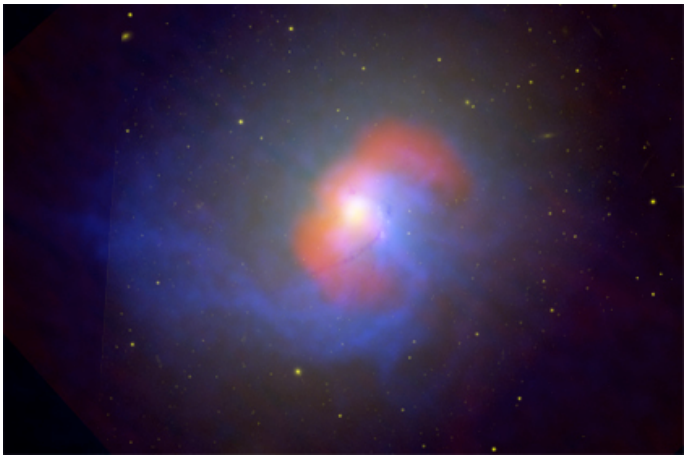

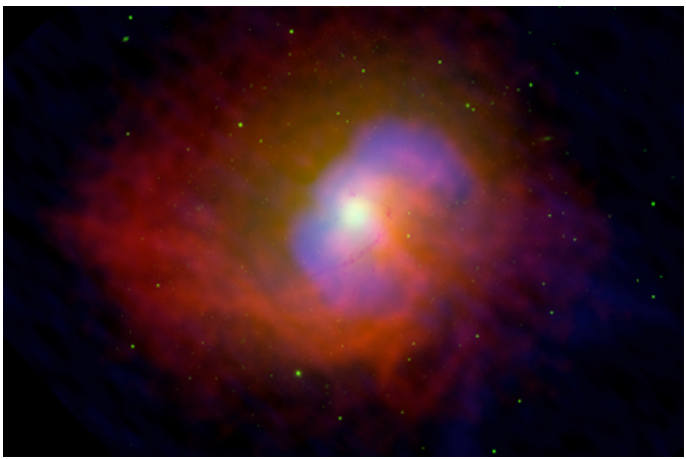

*Figure 1.* First item on survey, comparing attractiveness and temperature with the NGC 4696 X-ray/infrared/radio image.  Credit: X-ray: NASA/CXC/KIPAC/S.Allen et al; Radio: NRAO/VLA/G.Taylor; Infrared: NASA/ESA/McMaster Univ./W.Harris

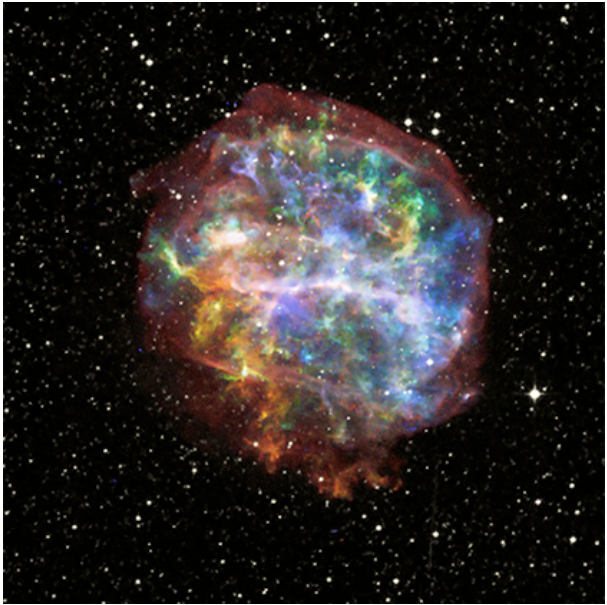

G292-A

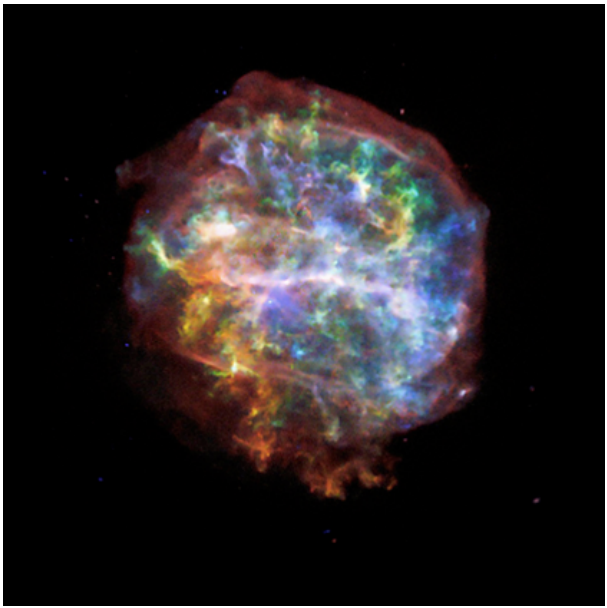

G292-B

G292.0+1.8 is a young supernova remnant located in our galaxy. This deep Chandra image shows a spectacularly detailed, rapidly expanding shell of gas that is 36 light years across and contains large amounts of oxygen, neon, magnesium, silicon and sulfur. Astronomers believe that this supernova remnant, one of only three in the Milky Way known to be rich in oxygen, was formed by the collapse and explosion of a massive star. Supernovas are of great interest because they are a primary source of the heavy elements believed to be necessary to form planets and life.

*Figure 2.* Images and text used for 2 (stars/no stars)*2(text/no text) item, showing young Galactic supernova remnant G292.0+1.8 in X-ray and optical light (A) and X-ray light only (B). Credit: X-ray: NASA/CXC/Penn State/S.Park et al.; Optical: Pal.Obs. DSS

The first dependent measure examined was the amount of time participants spent looking at the image (latencies). After truncating the most extreme 1% of the latencies, we conducted a two-way analysis of variance with image (stars or no stars) and text (explanatory text or no text) as independent variables. We found a main effect for text, $F(1, 8769) = 1373.69$, $p < .001$; but no effect for image, $F(1, 8769) = .104$, $p = .747$, or for the interaction term, $F(1, 8769) = 6.11$, $p = .013$. The means indicated that those participants reading the text spent longer on the image ($M = 33.27$, $SD = 22.86$) than those who did not receive the text ($M = 16.95$, $SD = 18.02$). The effect size is .80. There was wide variability in how much time participants spent in looking at the images, as can be seen from the standard deviations. Note that the mean amount of time spent looking at the image ranges from about 17 seconds with no text to 33 seconds with text. This time frame is the same order of magnitude as the results from the Smith and Smith (2001) study. In considering the differences here, it is apparent that the time difference (16.32 seconds) is roughly the amount of time an adult reader would take to read the 95 word explanatory text provided. Thus, it seems that the time spent looking at the image *per se* might be approximately the same in both conditions.

The next question asked how hot the image appeared to be, using a 10-point scale with 1 = Extremely cold, 5 = Neither hot nor cold, and 10 = Extremely hot. There was a main effect for image, $F(1, 8844) = 12.99$, $p < .001$, with the image without the stars being viewed as hotter ($M = 6.11$, $SD = 1.82$) than the image with the stars ($M = 5.98$, $SD = 1.82$), but the difference is rather small, with an effect size of .07. There was no main effect for text ($F(1, 8844) = 3.55$, *ns*) and no interaction term ($F(1, 8844) = 0.35$, *ns*).

Participants were also asked how attractive the image was, again using a 10 point scale, with 1 = Not at all attractive, 5 = Moderately attractive, and 10 = Very attractive. The results here were somewhat counterintuitive. There was no main effect for image ($F(1, 8844) = 2.12$, *ns*) nor for the interaction $F(1, 8844) = 0.01$, *ns*), but there was a main effect for text/no text, $F(1, 8844) = 79.57$, $p < .001$, an effect size of .19. When the image (either image) was presented with text, it was rated as more attractive ($M = 7.96$, $SD = 1.82$) than the image without the text ($M = 7.60$, $SD = 1.94$).

Next, two multiple-choice questions were asked of participants to examine the comprehension of the image they were looking at. The first question asked:

> What is this?
>  A.)  A galaxy
>  B.)  The leftover bits of an exploded star   *(*correct response)*
>  C.)  Gasses condensing around a black hole
>  D.)  An exploded planet
>  E.)  Don't know

Overall, 90.3% of participants chose the correct answer. In looking at differences between the two images and the text/no text conditions, there was a strong main effect for text condition, $F(1, 8752) = 314.71$, $p < .001$, a small, but significant main effect for image, $F(1, 8752) = 8.74$, $p = .003$, and a significant interaction term, $F(1, 8752) = 11.76$, $p < .001$. The main effect for text/no text showed that those participants receiving text were more likely to get the question correct (95.7%) than those not receiving the text (84.7%). As the text contained the answer, this is not surprising. The main effect for image and the interaction term can be understood when looking at them together. When the text is present, there are no differences between the images (image with stars = 95.6%, without stars = 95.9%). Without the text, the image with the stars is slightly more likely to engender a correct answer than the image without the stars (image with stars = 86.7%, without stars = 82.7%). Thus, the stars appear to provide a context that is helpful in interpretation in this instance.

The second question for this image asked:

How big across do you think this is?
A.) Smaller than the earth
B.) Between the size of the earth and the size of the solar system
C.) Between the size of the solar system and the size of the galaxy *(*correct response)*
D.) Larger than a galaxy

The text stated that the object is 36 light years across. Interestingly, the analysis of variance yielded no significant differences. Across all conditions, the percentages of participants answering correctly ranged from 71.0% to 73.0%. Thus, it appears to be the case that some participants in the text condition either did not process the information given or were not able to translate it into the comparative sizes of the solar system, galaxy, etc. Most wrong answers (21.1%) were, "Between the earth and the solar system."

*The Influence of Extensiveness and Type of Text.* The second section on the web-based survey/experiment focused on the influence of different approaches to providing textual information. The image for this experiment was Messier 51, also known as the Whirlpool Galaxy. This composite image is based on data from the Chandra, Spitzer, Hubble, and GALEX satellites (in X-ray, infrared, optical and ultraviolet light, respectively) and is presented in Figure 3.

As shown in Figure 3, participants were randomly assigned to one of four conditions in this experiment:

1. Image with no text (no text)

2. Image with text as it was originally written (original text)

3. Image with text written in a narrative style (narrative text)

4. Image with text with questions as headers (question-based text)

In the initial design of the study, there was a fifth condition, but it was dropped due to a typographical error in programming.

After viewing the image in one of the four experimental conditions, the participants were then asked a set of questions that were used, along with time spent on the image, as the dependent variables in a series of analyses of variance (ANOVA).

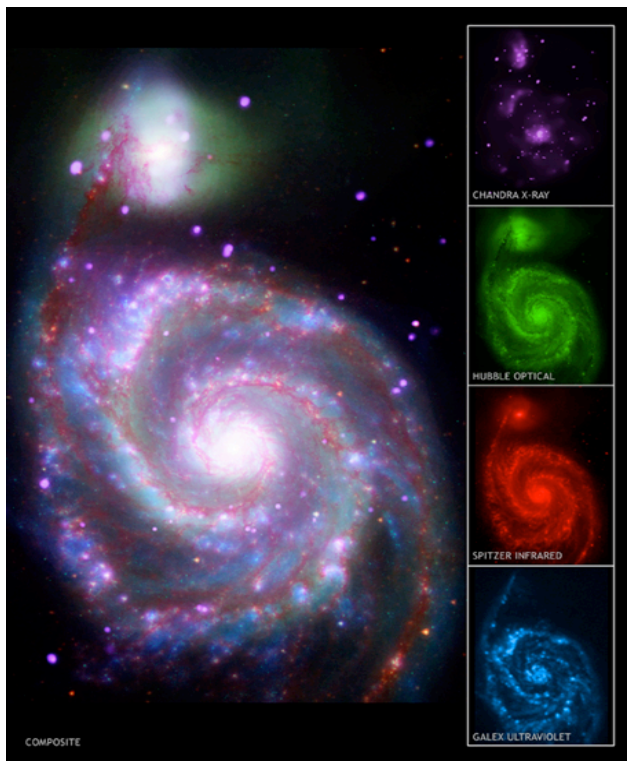

*Original Text*

A composite image of M51, also known as the Whirlpool Galaxy, shows a majestic spiral galaxy. Chandra finds point-like X-ray sources that are black holes and neutron stars in binary star systems, along with a diffuse glow of hot gas. Data from Hubble and Spitzer both highlight long lanes of stars and gas laced with dust. A view of M51 with GALEX shows hot, young stars that produce lots of ultraviolet energy.

*Text Written In Narrative Style*

Galaxies are born when filaments of gas collide. When these filaments are massive and they meet "head on," they create an *elliptical galaxy*; when they meet at an angle, they swirl and a *spiral galaxy* forms, such as M51, the Whirlpool Galaxy, depicted here. The sister galaxy, known as M51B, shown above the Whirlpool Galaxy is too small to be either elliptical or spiral; it is known as an *irregular galaxy*. This sister galaxy has passed through the Whirlpool Galaxy once and will over time splash into it again and become part of it. Like all big galaxies, the core of the Whirlpool Galaxy is bright because it is a dense collection of stars and gas with a giant black hole at the center.

The thumbnail images on the right were taken by the Chandra, Hubble, Spitzer, and GALEX satellites. The larger image on the left is a composite formed by adding the four images together. Each thumbnail depicts a different wavelength of light, generated by stars at different points in their life cycle. All colors are arbitrarily chosen by scientists.

*Text With Questions As Headers*

*Why does this look like a whirlpool?*
Galaxies are born when filaments of gas collide. When these filaments are massive and they meet "head on," they create an *elliptical galaxy*; when they meet at an angle, they swirl and a *spiral galaxy* forms, such as M51,  the Whirlpool Galaxy, depicted here.

*What is that other bright ball?*
The Whirlpool Galaxy has a sister galaxy, known as M51B, shown above it. This sister galaxy is too small to be either elliptical or spiral; it is known as an *irregular galaxy*. It has passed through the Whirlpool Galaxy once and will over time splash into it again  and become part of it.

*Why is this galaxy so bright in the middle?*
Like all big galaxies, the core of the Whirlpool Galaxy is bright because it is a dense collection of stars and gas with a giant black hole at the center.

*How was this picture made?*
The thumbnail images on the right were taken by the Chandra, Hubble, Spitzer, and GALEX satellites. The larger image on the left is a composite formed by adding the four images together. Each thumbnail depicts a different wavelength of light, generated by stars at different points in their life cycle. All colors are arbitrarily chosen by scientists.

*Figure 3.* Item with no text/text using varying presentation formats for Messier 51, also known

as the Whirlpool Galaxy.  Credit: X-ray: NASA/CXC/Wesleyan Univ./R.Kilgard et al; UV: NASA/JPL-Caltech; Optical: NASA/ESA/S. Beckwith & Hubble Heritage Team (STScI/AURA); IR: NASA/JPL-Caltech/ Univ. of AZ/R. Kennicutt

The first ANOVA used time as the dependent variable. Again, the most extreme 1% of the scores was removed from the data set. The ANOVA yielded a strong significant result, $F(3, 7077) = 571.25$, $p < .001$, *eta squared* = .20. The means and standard deviations are presented in Table 1. *Post hoc* analyses showed that the image alone ($M = 19.65$ sec., $SD = 22.29$) resulted in significantly less viewing time than the image with the original text ($M = 47.88$, $SD = 43.40$). The original text resulted in less viewing time than either of the longer texts. The narrative text ($M = 69.35$, $SD = 51.04$) engendered slightly less time than the question-based text ($M = 72.73$, $SD = 48.34$). (The latter two means were not significantly different.)  Of course, the texts differed in length, which may have accounted for the differences in viewing times. Using the mean latency of 19.67 seconds on the image with no text as a base, and a rough rule of thumb of a reading rate of 300 words per minute for this type of material, we estimated the additional viewing time of the image for each of the three text conditions. The initial text appears to have generated roughly 9 additional seconds of looking at the image (beyond the time estimated to read the additional text); the narrative text, 14 additional seconds; and the question-based text, 12 seconds. These are estimates, but it appears that the textual information caused participants to spend more time looking at the image, not simply reading the additional text.

Participants were asked, "In general, how well do you think you understand this image?" They responded using a 10 point scale with anchor points of 1 = Not at all, 5 = Moderately well, and 10 = Completely. There were strong differences among the various conditions, $F(3, 7053) = 156.34$, $p < .001$, *eta squared* = .06. Means and standard deviations are presented in Table 1. *Post hoc* analysis indicated that the image alone and the original text were not significantly different; whereas, both of the expanded versions resulted in higher ratings of perceived comprehension than either the image alone or the image with the original text.

Participants were then asked, "How much does the text help you to understand this image?" Again, a 10 point scale was used with anchor points of 1 = Not at all, 5 = Moderately, and 10 = Extremely. In this analysis, there were only three groups, as the group that did not receive any text was not included in the analysis. Differences by condition were statistically significant, $F(2, 5258) = 116.19$, $p < .001$, *eta squared* = .04. Means and standard deviations are presented in Table 1. *Post hoc* analyses showed that the narrative and question-based texts were rated more highly than the original text. These results are similar to the results on comprehension.

Table 1

*Descriptive Statistics for One-Way Analysis of Variance with Five Text Conditions*

| Variable | Text Condition | *n* | Mean | *SD* |
|---|---|---|---|---|
| Latency (sec.) for Image | Original Text | 1692 | 47.88 | 43.43 |
| | No Text | 1802 | 19.65 | 22.29 |
| | Narrative Text | 1779 | 69.35 | 51.40 |
| | Header Questions | 1808 | 72.73 | 50.48 |
| Comprehension (10-point scale) | Original Text | 1683 | 6.37 | 2.14 |
| | No Text | 1796 | 6.19 | 2.27 |
| | Narrative Text | 1777 | 7.27 | 1.94 |
| | Header Questions | 1801 | 7.39 | 1.88 |
| Helpfulness of Text (10-point scale) | Original Text | 1683 | 7.25 | 2.09 |
| | No Text | -- | -- | -- |
| | Narrative Text | 1777 | 8.01 | 1.76 |
| | Header Questions | 1801 | 8.15 | 1.70 |

*The Influence of Illustrative Scales and Expanded Text*

The next item on the survey concerned the presentation of illustrative scales and expanded text in the comprehension of an image from the Hinode Satellite showing spicules at the pole of the Sun. Participants were randomly assigned to one of eight conditions. There were four images: one with spicules only, one with a circle superimposed on the image that represented the diameter of the Earth, one with a circle that represented a three hundred mile diameter (the width of a typical spicule), and one with both the earth diameter circle and the three hundred mile diameter circle. The images were crossed with two text conditions: the original text for the image or an expanded text that explained in more detail what was in the original, shorter text. The four images and text formats are shown in Figure 4.

Two 4 (image) X 2 (text) analyses of variance were computed, one for the latencies and one for a score made from a 10 item true-false measure based on spicules. Once again, the most extreme 1% of the time scores were deleted from the data set. The latency analysis showed significant main effects for the scales ($F(3, 8773) = 98.52$, $p < .001$, *eta squared*) = .02 and for the text ($F(1, 8773) = 671.34$, $p < .001$, *eta squared* = .05); but no interaction ($F(3, 8773) = 1.76$, *ns*). The means and standard deviations are presented in Table 2.

The expanded text had an overall mean of 16.5 seconds more than the original text; given a difference of roughly 46 words, this time difference would suggest that participants spent slightly more time looking at the images than might simply be attributed to the amount of time needed to read the expanded text.

The means for the scales show that participants spent roughly 7 seconds more on the image with the Earth diameter circle than with no scales at all, and roughly 9 seconds more on the image with the 300 mile diameter circle than with the image with no scales. Participants spent roughly 16 seconds more on the image with both the 300 mile diameter circle and the Earth diameter circle than the image with no scale. This is the simple sum of the 7 second and 9 second differences. Thus, although it may be a coincidence, it is interesting to note that when presented with both scales, participants spent about as much time looking at each of the two scales as participants did when presented with either scale.

The second analysis was conducted on a score consisting of the total of correct responses that participants received on a 10-item true-false checklist of statements about spicules that they took after viewing the image. For this analysis, only the main effect for illustrative scale was significant, $F(3, 8773) = 36.21$, $p < .001$, *eta squared* = .01. A *post hoc* analysis showed that all three images with scales outperformed the image with no scale. Also, the image with both the Earth radius and the 300 mile radius circles led to higher scores than the image with the 300 mile radius circle only. Thus, the scales led to a better understanding of the image as compared to the image without any scale, although differences among the scales were not particularly strong.

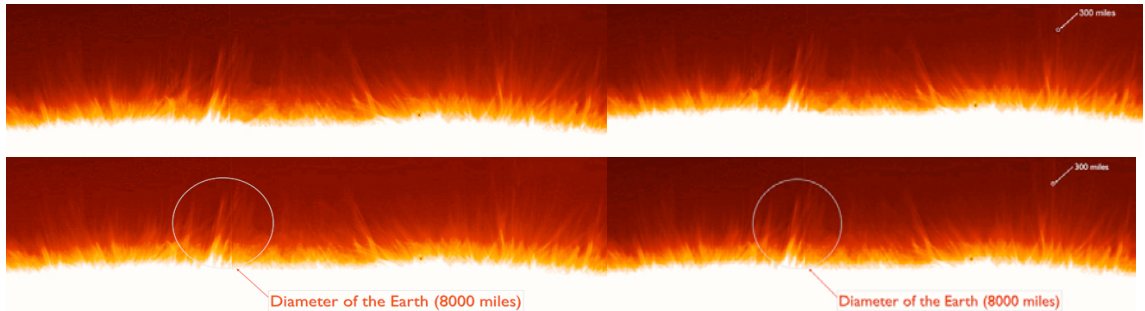

**Original Text**

The *Hinode* satellite has obtained high resolution images of the poles of the Sun. One of these, seen above, shows many resolved spicules jutting above the limb of the Sun, each of which is approximately 300 miles across.

**Elaborated Text – Both Circles**

The *Hinode* satellite has obtained high resolution images of the poles of the Sun. These are much like the North and South Poles of the Earth. One of these, seen above, shows many spicules, which are jets of hot gases, like filaments, that last for about 10 minutes. The widths of these spicules are resolved, meaning that the widths shown represent verified measurements. These spicules jut above the limb – or apparent edge – of the Sun, and are approximately 300 miles across.

To give an idea of scale, , the large superimposed circle indicates the diameter of the Earth (8000 miles across) and the small superimposed circle has a 300-mile diameter.

**Elaborated Text – Small Circle Only**

The *Hinode* satellite has obtained high resolution images of the poles of the Sun. These are much like the North and South Poles of the Earth. One of these, seen above, shows many spicules, which are jets of hot gases, like filaments, that last for about 10 minutes. The widths of these spicules are resolved, meaning that the widths shown represent verified measurements. These spicules jut above the limb – or apparent edge – of the Sun, and are approximately 300 miles across.

To give an idea of scale, the small superimposed circle has a 300-mile diameter.

**Elaborated Text – Large Circle Only**

The *Hinode* satellite has obtained high resolution images of the poles of the Sun. These are much like the North and South Poles of the Earth. One of these, seen above, shows many spicules, which are jets of hot gases, like filaments, that last for about 10 minutes. The widths of these spicules are resolved, meaning that the widths shown represent verified measurements. These spicules jut above the limb – or apparent edge – of the Sun, and are approximately 300 miles across.

To give an idea of scale, the large superimposed circle indicates the diameter of the Earth (8000 miles across).

*Figure 4.* Images and texts for 4(scale format)*2(text/no text) item.  Credit: X-ray: NASA/CfA/Hinode

Table 2

*Descriptive Statistics for Latency and Score Items on the Hinode Image*

| Hinode Image | Text Condition | *n* | Mean | *SD* |
|---|---|---|---|---|
| Latency Data (sec.) | | | | |
| 300 Mile Circle | Original Text | 1144 | 32.30 | 23.46 |
| | Elaborated Text | 1056 | 51.12 | 34.13 |
| Earth Circle | Original Text | 1148 | 31.65 | 26.78 |
| | Elaborated Text | 1040 | 47.92 | 32.21 |
| Both 300 Mile and Earth Circles | Original Text | 1115 | 40.29 | 31.68 |
| | Elaborated Text | 1108 | 55.70 | 33.86 |
| No Scale | Original Text | 1113 | 25.17 | 25.83 |
| | Elaborated Text | 1057 | 40.30 | 27.99 |
| ---------------------------------------------------------------------- | | | | |
| Score on Spicule Questions (10-point scale) | | | | |
| 300 Mile Circle | Original Text | 1144 | 8.00 | 1.31 |
| | Elaborated Text | 1056 | 7.97 | 1.26 |
| Earth Circle | Original Text | 1148 | 8.10 | 1.25 |
| | Elaborated Text | 1040 | 8.00 | 1.24 |
| Both 300 Mile and Earth Circles | Original Text | 1115 | 8.15 | 1.23 |
| | Elaborated Text | 1108 | 8.10 | 1.25 |
| No Scale | Original Text | 1113 | 7.79 | 1.30 |
| | Elaborated Text | 1057 | 7.71 | 1.23 |

*Experts Novices Differences in Perception of Astronomical Images*

The second research question concerned how experts differ from novices in their perception of astronomical images. We used both the data from the web-based survey/experiment and the focus group data to address this question. In the web-based survey/experiment, we asked participants to rate their knowledge of astronomy on a 1 = novice to 10 = expert scale. We screened responses to filter out highly unlikely combinations, such as participants who rated themselves as experts, but who had little formal education, or were very young. Although such individuals may exist, they are much less likely to occur than individuals who are older and have higher levels of education. We checked filtered individuals against scores on the spicules measure described above, which confirmed the screening process. The distribution of self-ratings of expertise is presented in Figure 5. As can be seen, the majority of the participants rated themselves between a 3 and an 8 in terms of expertise. However, the large sample size in the study allowed for comparisons of experts (n = 478) and novices (n = 1376).

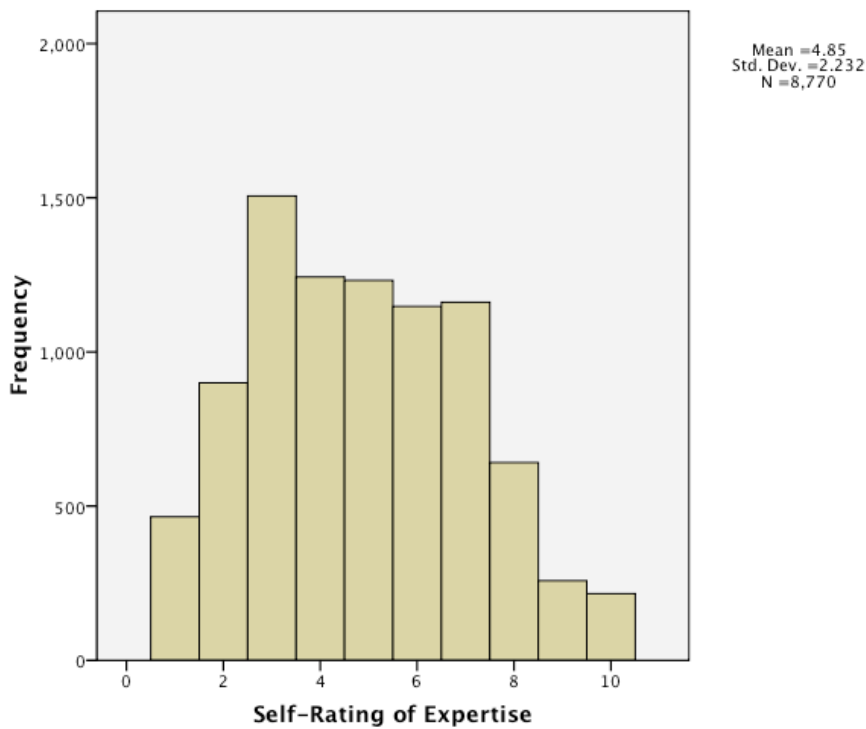

*Figure 5.* Distribution of self-ratings of expertise.

The first analysis looked at the "red" and "blue" versions of the NGC 4696 X-ray/infrared/radio image. A contingency table *chi* square analysis was performed on which image participants preferred and which image they thought was hotter in temperature. For the preference question, more novices thought the "blue" version was more attractive (56.6%), whereas the experts thought the "red" image was more attractive (55.3%), $X^2$(1, N = 1836) = 19.96, *p* < .001. For the "which is hotter" question, the novices thought the red image was hotter by 77.6% to 22.4%, whereas the experts thought the red image was hotter by 62.0% to 38.0%; $X^2$(1, N = 1836) = 43.92, *p* < .001.

The second analysis examined latencies for expert/novice differences using the image with and without stars, and with and without explanatory text. There was a significant interaction effect, *F*(1, 1823) = 31.91, *p* < .001, *eta squared* = .02. Without the text, the mean viewing times between experts and novices were fairly similar, with M(novices) = 17.32, *SD* = 17.57 and M(experts) = 14.79, *SD* = 16.00. However, when the text was present, the differences were substantial, with M(novices) = 37.15, *SD* = 22.74 and M(experts) = 23.00, *SD* = 16.84. Novices spend approximately 20 seconds more on the image when the text was provided, versus approximately 5 seconds more for experts. The main effect for expert/novice differences overall was also significant (*F*(1, 1823) = 65.20, *p* < .001, *eta squared* = .04), but not important in light of the interaction effect.

We next looked at the responses to the questions about what the object was and how large it was. As would be expected, there were significant main effects for expert/novice differences for identifying the object (*F*(1, 1819) = 121.41, *p* < .001, *eta squared* = .06); and its size (*F*(1, 1819) = 113.09, *p* < .001, *eta squared* = .06). For the "what is this" question, there was also an interaction between expert/novice differences and the presence or absence of explanatory text, $F(1, 1819) = 41.08$, $p < .001$, *eta squared* = .02. Without the explanatory text, mean differences in performance were quite large, with 95.2% of experts answering the question correctly, as compared to 61.4% of the novices. With the explanatory text, the differences were far less, with 97.8% of experts answering the question correctly, as compared to 88.8% of the novices. The question concerning how large the image was produced a strong expert/novice difference with no interaction with the image or the text. Experts correctly answered the question 83.1% of the time, as compared to 56.7% for novices.

The next analysis examined the image of the Whirlpool Galaxy, M51, with the four text conditions to determine the interaction between expert/novice differences and how well participants felt they understood the image (perceived comprehension), and how helpful they thought the text was. For the perceived comprehension question, the analysis was significant for both main effects and for the interaction term. Means and standard deviations are presented in Table 3. The interaction, $F(3, 1460) = 5.82$, $p < .001$, *eta squared* = .01, revealed that the overall pattern of responses was similar for experts and novices, but that novices reported comparatively less comprehension than did their expert counterparts when not provided with any text, or when provided with the basic original text. The main effect for text was $F(1, 1460) = 24.91$, $p < .001$, *eta squared* = .05. A *post hoc* analysis indicated that the question-based text resulted in the highest level of perceived comprehension, followed by the narrative text, the original text, and no text. The expert/novice difference was strong, as would be expected, with an $F(1, 1460) = 945.41$, $p < .001$, an effect size of 2.01 between experts and novices, with novices being much more dependent on the explanatory text.

When asked how helpful the text was to their comprehension, a somewhat different picture emerged. Again, both main effects and the interaction term were significant (see Table 3). The main effect for expert/novice was $F(1, 1108) = 5.89$, $p = .015$. The interaction effect was $F(3, 1108) = 9.61$, $p < .001$, *eta squared* = .02, and showed that the novices that felt the question-based text was the most helpful and the original text the least helpful, with the narrative text closer to the question-based text in helpfulness. The experts, on the other hand, found all three texts somewhat similar. It appears that the experts do not need nor particularly desire the text to be put in a narrative form or to have questions serving as an organizational device; they just want the information.

The differences between experts and novices seen here led us to wonder what the results (in terms of perceived comprehension and helpfulness) would look like for all participants. We constructed two graphs to examine this issue (Figures 6 and 7). The first graph shows the perceived comprehension of the image by self-rating of expertise. Here, we see a fairly linear increase by expertise, with the original text and no text groups reporting the least comprehension overall, and the two expanded texts showing higher levels of perceived comprehension. The graph for helpfulness shows a somewhat different pattern. (Recall that the "no text" group is not included here). What is shown here is that in terms of helpfulness, the expanded texts seem to be most helpful to people with self-ratings of expertise between about 4 and 8, with the expert group finding these texts less helpful than those with self-ratings of 4 to 8. The original text is most helpful for people with self ratings of 8 and 9, again dropping off at the extremes. Thus, it seems that the expanded texts seem to be targeted at the middle to high range of expertise, and the original text at a higher level of expertise. Also, people who rated themselves as expert tend to find all texts somewhat less helpful than those who are less expert.

Table 3

*Descriptive Statistics for Latency and Score Items for M51 (Whirlpool Galaxy)*

| Variable | Expert/Novice | Text Condition | *n* | Mean | *SD* |
|---|---|---|---|---|---|
| Comprehension | | | | | |
| | Novice | Original Text | 261 | 4.45 | 2.16 |
| | | No Text | 272 | 4.21 | 2.25 |
| | | Narrative Text | 263 | 5.73 | 2.17 |
| | | Header Questions | 277 | 5.93 | 2.15 |
| | Expert | Original Text | 101 | 8.50 | 1.51 |
| | | No Text | 82 | 8.39 | 1.80 |
| | | Narrative Text | 108 | 8.88 | 1.01 |
| | | Header Questions | 104 | 9.04 | 1.04 |
| Helpfulness of Text | | | | | |
| | Novice | Original Text | 261 | 6.36 | 2.53 |
| | | Narrative Text | 263 | 7.57 | 2.09 |
| | | Header Questions | 277 | 7.96 | 2.04 |
| | Expert | Original Text | 101 | 7.59 | 2.01 |
| | | Narrative Text | 108 | 7.45 | 1.92 |
| | | Header Questions | 104 | 7.87 | 1.90 |

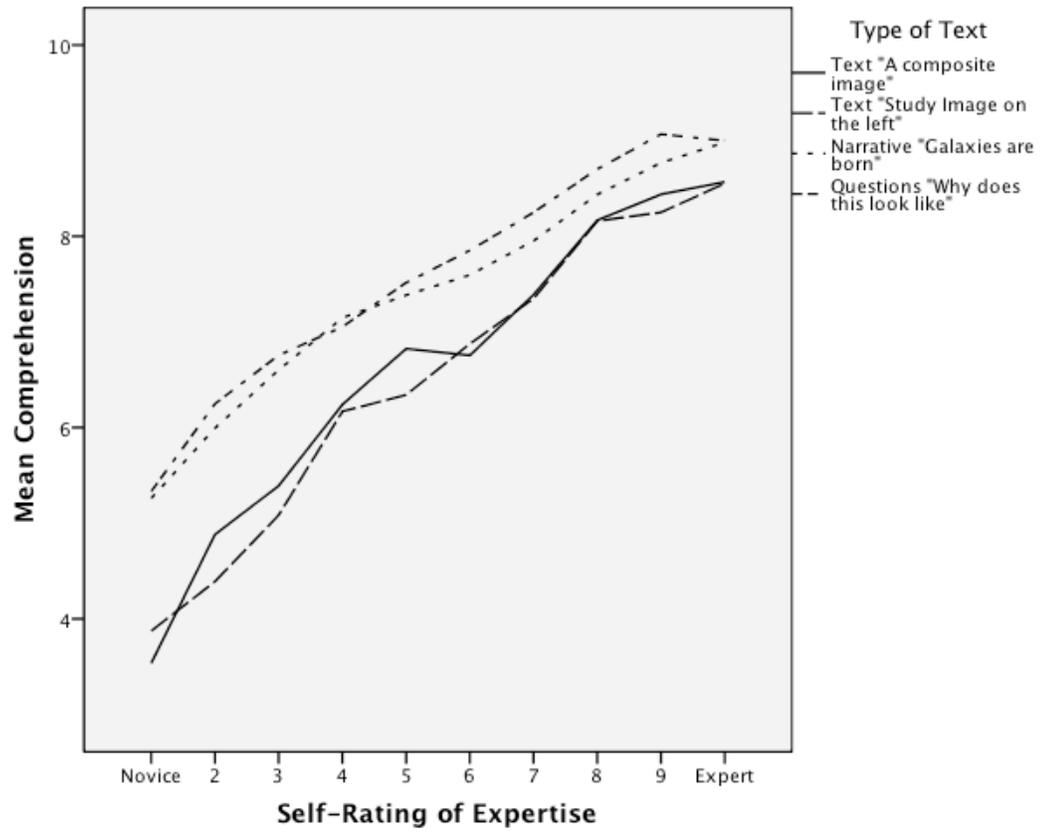

*Figure 6.* Graph of comprehension of text by all levels of self-reported expertise.

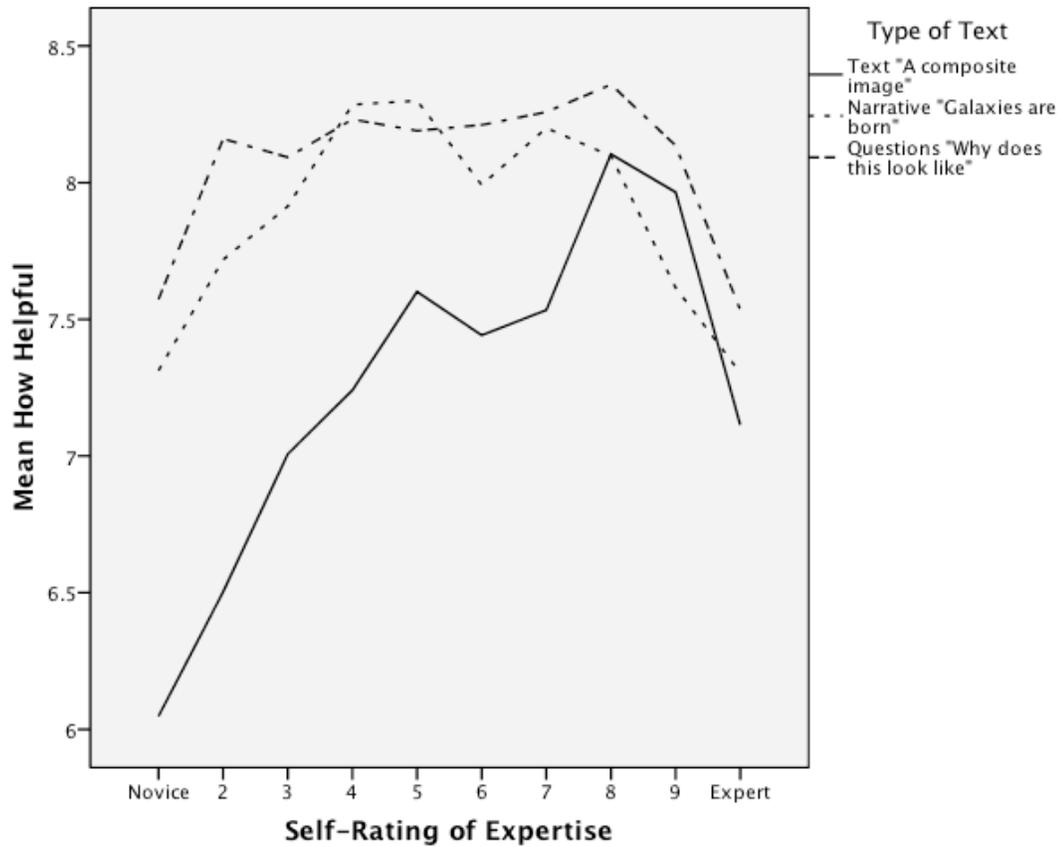

*Figure 7.* Graph of helpfulness of text by all levels of self-reported expertise.

The third analysis involving expert/novice differences concerned the Hinode images and texts. Here, we looked at the type of scale presented (or the absence of any scale), short or expanded explanatory text, and expert/novice differences. The dependent variables were time spent looking at the image, and the score on the 10 item true/false measure on spicules. For both dependent measures, each of the main effects were significant, but none of the interactions. The main effects for illustrative scale and for text have been discussed above. For the latency data, the results, $F(1, 1822) = 86.38$, $p < .001$, Cohen's $d = .46$, showed that experts spent far less time looking at the image ($M = 30.32$, $SD = 28.03$) than did novices ($M = 44.46$, $SD = 32.82$). Their comprehension though, was stronger, $F(1, 1822) = 185.77$, $p < .001$, Cohen's $d = .71$, with M(experts) = 8.55, $SD = 1.31$ and M(novices) = 7.75, $SD = 1.25$.

*The Focus Group Study*

We also examined expert/novice differences through a series of focus groups conducted at the Harvard/Smithsonian Center for Astrophysics (CfA). A number of topics and issues were explored in the focus groups. For this analysis, we focus on how experts and non-experts approached looking at astronomical images.

To begin, the three groups of novices in the focus groups (students, teachers, and lay public) were probably more knowledgeable about astronomy than typical students, teachers, or members of the public in that they responded to advertising about participating in such a focus group at the CfA. Hence, we refer to them as "non-professionals" as opposed to novices. With that as a caveat, and acknowledging that this may raise issues regarding the ethics of representation, we found what we felt were useful and important differences among the groups.

Each focus group was shown two images and asked a series of questions to ascertain their knowledge about, misconceptions of, interest in learning about, and aesthetic reactions to each image. The first image was M101, which had not yet been released, pending the launch of NASA's International Year of Astronomy activities in January 2009. The second image was of the spicules on the pole of the Sun, with two scales shown: a circle representing the diameter of the Earth and a circle representing the diameter of an average spicule (see Appendix A). This was the same image shown the participants in one of the groups in the web-based survey/experiment, but without the accompanying text. Recall that none of the focus group participants had participated in the web-based survey/experiment.

The images were projected onto a large screen in a darkened room, thus producing a rather dramatic image when the image first appeared. The non-professionals responded to the images with what might be called a sense of awe or wonder. There were typically "oohs" and "ahs" in the room when the images first were projected. The professionals, on the other hand, typically initially reacted with a sense of recognition at the object. There were "ahs" in this group as well, but they were "ahs" of "Ah, this is the Hinode image of the pole of the Sun," as opposed to "Ah, this is amazing."  It should be noted that the non-professionals also had a good idea of what they were looking at (clearly a spiral galaxy and an image of the Sun), but not with the level of specification of recognition of the professionals.

In the discussions that ensued from the projection of the images, the non-professionals wanted to know what the colors represented, how the images were made, whether the images were composites from different satellites, and what various areas of the images were. They wanted to know if M101 could be seen with a home telescope, binoculars, or the naked eye. They commented that the longer they looked at the objects, the more questions they had. They wanted to know how much manipulation of the image was involved in its creation. They felt that the use of scales was helpful. They stated that explanations of images can be technical, but that there should be explanations in lay language, and that they should include the historical context, discovery background, what scientists find interesting about the image, and what scientists feel is important to observe in the image, perhaps as "rollovers" on the computer screen. Several of the non-professionals indicated that they wanted to be able to observe the image more in the way that scientists do. To summarize non-professional reactions to the images, they began with a sense of awe, and then moved toward a more inquisitive stance wanting to know more about the generation of the image, next to the science behind the image, and finally to how scientists would look at the image (a sort of, "What should I be noticing that I am not?" perspective).

The astrophysicists, by comparison, did not appear to be awed by the initial presentation of the images. Even if they had not seen this particular representation of the galaxy or the Sun, they were certainly familiar with the object and with other representations of it, if not the one currently on the screen. They seemed to be taking in information about a familiar object as opposed to being in awe of something fundamentally beautiful. Their first reactions were typified by the comment, "I want to know who made this image and what it was that they were trying to convey. I want to judge whether this image is doing a good job of telling me what it is they wanted me to get out of this."  Still, when asked directly, most of the astrophysicists said that they felt they engaged in a basically aesthetic as opposed to scientific activity when looking at the images. However, as one of the professionals explained, "…you'd have to be a moth not to be impressed with the aestheticism of it…a lot of these astronomical images are just outright beautiful."

The astrophysicists took a critical eye to the images. Interestingly, they were very much concerned with how the images conveyed information to the general public. Several thought that the M101 image was "too busy." They were concerned that the public learns the science behind the image, why it was important, where it was in the sky, and the scale of the object. They felt the image and accompanying information should talk about the wavelengths in the image, what the colors represent, the layers of the image, and when it was created.

It is difficult to generalize too far from the results of the focus groups, but some tentative ideas about differences in viewing astronomical images between experts and novices taken from the focus group study might be:

- A tendency for novices to work from aesthetics to science, and for astrophysicists to work from science to aesthetics.

- Astrophysicists look toward the scientific communication of the image—why has the image been created and what is being communicated by it, whereas novices are more oriented toward learning more about the image and the science underlying it.

- Both groups are interested in issues of just what is being presented.

- Novices want to learn how to look at the images the way the scientists do.

## Summary and Discussion

Our goal in this research was to take an initial look at how people view astronomical images, how they process the information from those images, and how augmentation in terms of text and scales influences their perception. We additionally looked at expert/novice differences. We found that augmentation is critical for appreciation of the images, especially for those who are not experts, and that experts and novices differ substantially in how they look at these objects. What does this tell us in terms of scientific communication in general?  To begin, astronomical images are highly captivating to a wide variety of audiences; they engender a sense of awe and wonder. Thus, communication about astronomical images begins from a different starting point than many other forms of scientific communication. We have their attention, but we also have a story to tell that can rapidly get highly technical and difficult to comprehend. One of the major findings of this study is that language that is conversational and engaging in nature, that anticipates an engaged, intelligent, but not sophisticated audience, seems to be the best way to provide information, especially to non-experts. This is very similar to what one would find in communicating about fine art (see e.g., Smith, & Wolf, 1996). At some level, this is an excellent situation for science educators find themselves in: reaching out to people who are eager to hear an interesting story that enhances their understanding of the scientific topic at hand. Using the results obtained here may best deliver that story in a way that satisfies the immediate issues and questions, and also invites the viewer to discover more information.

Overall, the results point toward a viable way to answer Wyatt's (2008) call for making astronomical images "more meaningful and accessible" (p. 26). In addition to the findings regarding text, illustrative scales that provide a sense of the size of an astronomical image were useful to novices and experts alike, although we did not find strong differences among the type of scales used. The use of color in the images appears to greatly enhance their attractiveness, but is clear that individuals do not often (or even typically) interpret color in the same fashion as the creator of the image intended it. We note that the time spent looking at images, and the tendency, especially for novices, to read the accompanying textual material is similar to what is found in studies of individuals looking at art (Cupchik, Shereck, & Speigel, 1994; Smith & Smith, 2001).

Although experts appreciate the aesthetic value and sense of wonder conveyed in astronomical images, their initial approach to images appears to be one of "what is the information that is being communicated in this image?" They typically know what the object is and what its characteristics are, and so look for what the creators of the image were trying to convey. They want to know what wavelengths the colors represent, and how this image differs from other similar images, perhaps even of the same object. Interestingly, the experts in this study spent less time on average looking at the images as compared to the novices. They also spent less time reading the text that accompanied the images. This may be due to having stronger background knowledge, and therefore being able to read the text more quickly, without having to actively cognitively process terms that are unfamiliar to novices. The experts in this study also were more likely to prefer shorter texts and texts not written in a narrative fashion.

Novices, on the other hand, appear to have initial reactions to astronomical images of awe and wonder. They are taken aback by the beauty, and perhaps the mystery, of the images. They begin with the sense of amazement, and then begin to form questions in their minds about the images: what and where is what they are viewing, how big and how far away is the object, how was the image made? In comparison to the experts, the novices in this study preferred more information and information that had been contextualized or written in a fashion that deliberately attempted to facilitate their comprehension. They spent more time looking at the images, and more time reading the text that accompanied them. They wanted to learn how to look at the images in the way that astrophysicists do. In this sense, they are like their counterparts looking at art. They take the image in for what it is and do not immediately classify the image as representing something specific – whether a particular type of wavelength for astronomical images or representing an early or late work of an artist.

The findings from this research are based on two data sources, each with limitations. The survey/experiment was based on individuals who were invited to participate in the study primarily from astronomy web sites, and they were volunteers. Thus, although being able to randomly assign participants to conditions is a clear strength of the study, it must be acknowledged that these were mostly people with an extant interest in astronomy. The utility of instructional devices such as explanatory text and scales might be even stronger with a sample of participants with less of an orientation or positive affect toward astronomy. On the other hand, the findings are probably readily generalizable to those individuals who are interested in astronomy. For the focus group study, the strength was the ability to gather a group of professional astronomers to look at issues of astronomical images and how their science is communicated. The weakness in the focus group results is that the "novice" groups were probably substantially more sophisticated than the labels of "lay public," "students," and "teachers" might suggest. It is our intent to replicate these latter groups with individuals less keenly interested in astronomy. We also recognize that our measures are beginning efforts, but we feel we have learned much from them, and plan to move on to the development of instruments that might be used across a variety of settings.

*Applying the Findings*

Within a few months of participating in the focus groups, the Education and Public Outreach (EPO) group for the Chandra X-ray Observatory Center began work on implementing some of the specific shared comments, particularly those expressed from the focus group of students. Some of these applications on the Chandra web site were relatively straight-forward, such as including bulleted text for overview and quick comprehension in the body of new images released, interactive labeling on the images to better show what the scientist wants the viewer to see, and providing more "Wikipedia-style" links (as the students termed it) for the body of the text of new images. The second step was developing a simple interactive multiwavelength image feature that allows the user to move from one color or wavelength to the next to "build" the complete image composite to "see" how it was made. This website can be accessed at http://chandra.harvard.edu/photo/2009/jkcs041/. Since implementing this change, the Chandra EPO group has received a high percentage of positive comments, in particular on the multiwavelength feature. The EPO group is implementing a brief survey questionnaire to collect visitor data specific to the multiwavelength feature and interactive labeling to determine the impact of the features on the public's comprehension.

More recently, the EPO group has built an interactive question-based text script into its newly released photo pages with click-tracking methods to count the user clicks per question, per image, and compare totals.  They have, similarly, implemented question-based text into a series of print products. These new poster-sized give-aways for the public feature multiwavelength astronomical images, with the individual color filters provided as additional images, and question-based text that explores the "Who, What, Where, When, and How" of the science behind the image. The text highlights some of the content that was commonly asked during the focus groups including how the images were made, the historical importance of the object, the location in the night sky, etc. Data collection and a brief summative evaluation of these six posters are being conducted to analyze the impact of the improved features on the public's understanding.

Returning to other possible recommendations from the findings, it might be useful to provide color scales, along with physical scales, in images intended for the public. On images that involve multiple wavelengths, a color key could be provided that denotes what wavelengths are associated with each color. The finding that viewers responded well to two physical scales suggests that a physical scale and a color scale might be provided without a loss of interest to the viewer.

Another intriguing finding is that many novices want to understand how experts view the images. It would be possible to provide such information with images in the future, to have a "rollover" on the image that says, "Here is what astronomers see….", or have commentary from astronomers available.

*Looking Ahead*

The A & A group has a study in progress that applies the findings from this research in four major science museums in the United States and one in New Zealand, to determine if similar results are obtained in situ. Additionally, we will be looking more carefully at differences in color schemes and their communicative efficacy and looking at more interactive approaches to communicating the science that underlies the astronomical images, including the use of mobile devices.

We are also interested in looking at the effects of viewing astronomical images. How does viewing such images cause people to view science differently? Does looking at astronomical images cause children to look more positively at considering science as a potential career? Can we extend the parallels between looking at astronomical images and looking at art? We feel we need to continue to explore these questions both using the techniques we have presented here, but also need to find communities less inherently interested in science than our current samples to explore the impact of astronomical images on a broader public.

Acknowledgements

This project was developed with funding from the Hinode X-ray Telescope, performed under NASA contract  NNM07AB07C, as well as Education and Outreach group for NASA's Chandra X-ray Observatory, operated by SAO under NASA Contract NAS8-03060. The authors thank Dr. Jerry Bonnell (NASA's GSFC, CRESST/USRA) and APOD, without whom the high response rate to the web-based survey would not have been possible.

Appendix A
Questions for Semi-Structured Interview for Focus Groups

1. How many of you have seen this image before?
2. Those of you who have seen this image before, do you recall where?
3. How do you go about looking at the image? What do you look at first?
4. What came to mind when you first saw this image? Everyone think of an answer here individually and then we'll quickly go around the room to see what folks say.
5. What questions do you have when you look at this image?
6. How much are you interested in what this is?
7. What do you think caused it?
8. How big do you think it is? (prompt with relative to the size of the Earth if needed)
9. How stable do you think is it?
10. How solid is it?
11. How distant is this from the Earth?
12. Is this a common occurrence in space or is this unique?
13. How much do these colours represent what it would actually look like if you could see it actually out in space?
14. What colours do you view as hot? Cold?
15. How much do you want to know about the scientists who captured this image? Things like, where they are from? What sort of work do they do? What satellite was used to grab this image? How did they get this kind of job?
16. If the astronomer who specialised in the area of this image were here, what would ask him/her?
17. What kinds of information would you like that would enhance your viewing of object like this one?
18. What sorts of written explanations do you look for? Do you like technical explanations? Lay language? Should technical terms be defined? Do you look for links where you can learn more?
19. What are your favourite areas in astronomy?
20. What would you like to learn more about?
21. How much interest do you have in how to have a career like the person who grabbed this image?

First Image: M101

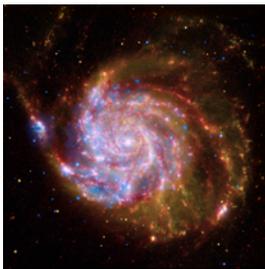

Second Image: Spicules on the Pole of the Sun (with scales to aid comprehension)

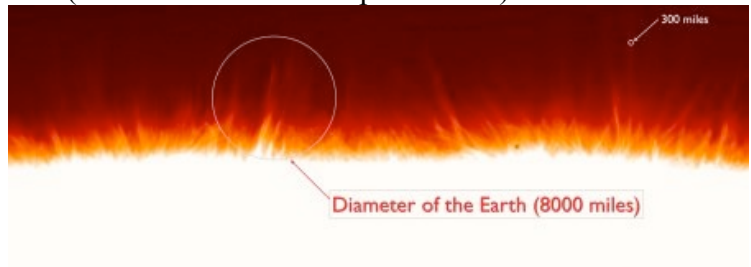

Credit: X-ray: NASA/CXC/JHU/K.Kuntz et al.; Optical: NASA/ESA/STScI; IR: NASA/JPL-Caltech/STScI/K. Gordon

Credit: NASA/CfA/Hinode

Appendix B
Focus Group Paper-and-Pencil Activity

*Participants responded to the following set of items for each picture shown:*
1. What is it?
2. How big is it in relationship to the Earth?
3. How old is it?
4. What is it made of?
5. Is this common or unique in the universe?
6. How much does this image represent what you'd actually see if you were out in space?

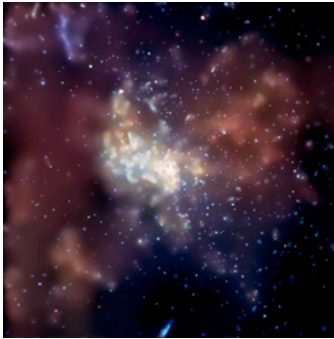

Sagittarius A* (region around the black
hole at the center of our galaxy)
Credit: NASA/CXC/MIT/F.K.Baganoff et al.

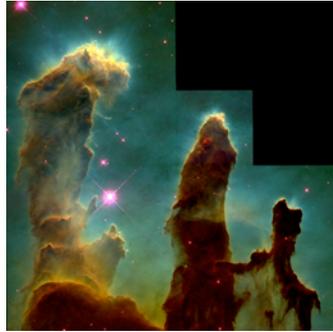

The Eagle Nebula (also known as
Messier Object 16 or NGC 6611)
Credit: NASA/ESA/STScI/ASU/J.Hester & P.Scowen.

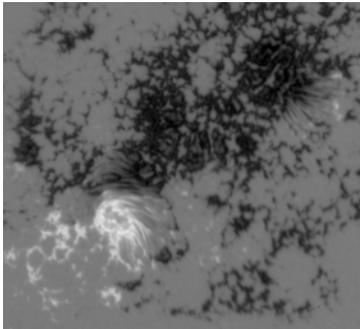

A Sunspot (observed by Hinode's
SOT Narrowband Filter Imager)
Credit: NASA/CfA/Hinode

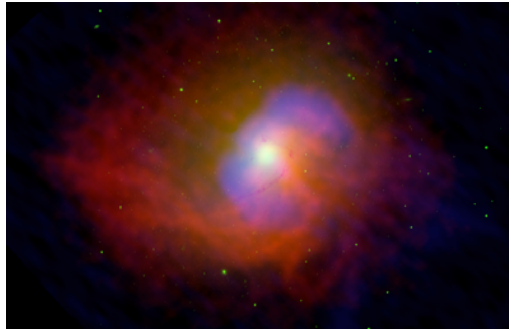

NGC 4696 (a large elliptical galaxy
in the Centaurus Galaxy Cluster)
Credit: X-ray: NASA/CXC/KIPAC/S.Allen et al; Radio:
NRAO/VLA/G.Taylor; Infrared: NASA/ESA/McMaster
Univ./W.Harris

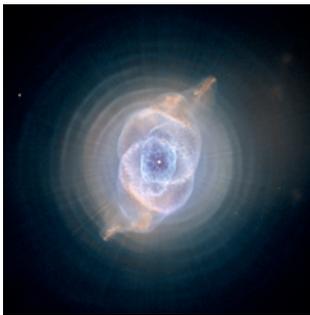

Cat's Eye Nebula or NGC 6543
Credit: NASA/STScI